\documentclass[aps,prl,twocolumn,showpacs,preprintnumbers,superscriptaddress]{revtex4}
\usepackage{amssymb}
\usepackage{graphicx}
\usepackage{dcolumn}
\usepackage{bm}

\newcommand{\nc}{\newcommand}
\nc{\be}{\begin{equation}}
\nc{\ee}{\end{equation}}
\nc{\bea}{\begin{eqnarray}}
\nc{\eea}{\end{eqnarray}}
\nc{\bean}{\begin{eqnarray*}}
\nc{\eean}{\end{eqnarray*}}
\nc{\mb}{\mbox}
\nc{\rnc}{\renewcommand}
\nc{\vk}{\mb{\bf k}}
\nc{\vp}{\mb{\bf p}}
\nc{\vn}{\mb{\bf n}}
\nc{\vq}{\mb{\bf q}}
\nc{\rr}{\mb{\bf r}}
\nc{\vz}{\hat {\mb{\bf z}}}
\nc{\vj}{\mb{\boldmath$j$}}
\nc{\vg}{\mb{\boldmath$g$}}
\nc{\x}{\mb{\boldmath$x$}}
\nc{\A}{\mb{\boldmath$A$}}
\nc{\va}{\mb{\boldmath$a$}}
\nc{\vs}{\mb{\boldmath$\sigma$}}
\nc{\vpi}{\mb{\boldmath$\pi$}}
\nc{\nab}{\nabla}
\nc{\X}{\sf x}

\begin{document}

\title{Anomalous Exciton-Condensation in Graphene Bilayers}
\author{Yafis Barlas}
\affiliation{National High Magnetic Field Laboratory, Florida State University,
Tallahassee, FL, 32306,USA}
\author{R. C\^{o}t\'{e}}
\affiliation{D\'{e}partement de Physique, Universit\'{e} de Sherbrooke, Sherbrooke, Qu%
\'{e}bec, Canada, J1K 2R1}
\author{J. Lambert}
\affiliation{D\'{e}partement de Physique, Universit\'{e} de Sherbrooke, Sherbrooke, Qu%
\'{e}bec, Canada, J1K 2R1}
\author{A.H. MacDonald}
\affiliation{Department of Physics, The University of Texas at Austin, Austin Texas 78712}

\begin{abstract}
In ordinary semiconductor bilayers, exciton condensates appear at total Landau 
level filling factor $\nu_{T}=1$.  We predict that 
similar states will occur in Bernal stacked graphene bilayers at many non-zero integer filling factors. 
For $\nu_{T} = -3,1$ we find that the superfluid density of the exciton condensate 
vanishes and that a finite-temperature fluctuation induced first order
isotropic-smectic phase transition occurs when the layer densities are not balanced. 
These anomalous properties of bilayer graphene exciton condensates 
are due to the degeneracy of Landau levels with $n=0$ and $n=1$ orbital character.
\end{abstract}

\maketitle

\noindent \emph{Introduction}--- When two semiconductor quantum wells occupied by 
half-filled Landau levels are narrowly separated, the
bilayer system ground state spontaneously establishes
inter-layer coherence \cite{coherencerefs}.  These broken symmetry states 
possess a condensate of pairs (each composed of an electron in
one layer and a hole in the other) which opens up an energy gap
responsible for a total filling factor $\nu _{T}=1$ quantum Hall
effect, supports a
dissipationless counterflow (excitonic) supercurrent, and is responsible for a wide variety of
incompletely understood transport anomalies \cite{transportrefs,naturereview}.
Recently an interesting new type of bilayer two-dimensional electron
system has become available \cite{graphenebilayerexpt,graphenermp} which consists of two carbon-atom
honeycomb-lattice (graphene) layers separated by a fraction of a nanometer.
The electronic structure of graphene bilayers is full of surprises \cite%
{graphenebilayertheory} because of an interplay between the sublattice
pseudospin chirality of each layer \cite{chiral2degprl} and the Bernal 
stacking arrangement, particularly so in the presence of an external
magnetic field. In this structure, one of the two-carbon atom sites in each
layer has a near-neighbor in the other layer and one does not. Interlayer
hopping drives electrons on the closely spaced sites away from the Fermi
level, leaving one low-energy site for carbon $\pi $-orbitals in each layer.
Because the hopping process between low-energy sites occurs in two steps,
via an intermediate high-energy state, it turns out\cite%
{graphenebilayertheory} that in the presence of a magnetic field both $n=0$
and $n=1$ Landau level orbitals have zero kinetic energy, and that the corresponding 
wavefunctions are localized in opposite layers for opposite ($K$ and $K^{\prime }$) graphene
valleys \cite{graphenermp}.  Even though
the two-layers are close together and band eigenstates at zero-field are a
coherent combinations of individual layer components, the zero-energy strong
field states occupy a definite layer.

In the absence of electron-electron and Zeeman interactions, a group of eight
degenerate
Landau levels (the bilayer octet) develops at the
Fermi level of a neutral graphene bilayer because of the degeneracy between
$n=0$ and $n=1$ states combined with spin and valley degeneracy.  The octet is revealed by a jump
in the quantized Hall conductivity \cite{graphenebilayerexpt} from $%
-4(e^{2}/h)$ to $4(e^{2}/h)$ when the charge density is tuned across
neutrality in moderately disordered samples. When the sample quality is
sufficient for interactions to dominate over disorder, quantum Hall effects
occur \cite{kostya2007,ourprl,octetexperiment} at all intermediate integers.
Spontaneous valley-symmetry breaking is expected\cite{ourprl} in all but the $\nu_{T}=0$ (half-filled octet) case.
Interactions favor states with spontaneous coherence between valleys, 
and hence between layers, because this broken symmetry does not require 
charge transfer between layers.  In this Letter we
illustrate the unique and rich properties of bilayer graphene exciton condensates
by concentrating on the simplest case in which the octet is occupied either by single majority-spin
($\nu _{T}=-3$) or a single minority-spin ($\nu _{T}=1$) Landau level, and allowing 
for an external potential difference $\Delta _{V}$
between the layers.  We demonstrate that the 
superfluid density vanishes at these filling factors, and that a finite-temperature fluctuation induced first order
isotropic-smectic phase transition occurs when the layer densities are not balanced. 
These anomalous properties of bilayer graphene exciton condensates 
are due to the degeneracy of Landau levels with $n=0$ and $n=1$ orbital character.
Below we first discuss the mean-field ground state,
demonstrating that it exhibits spontaneous inter-layer phase coherence for
small $\Delta _{V}$ at the filling factors of interest. We then explain why
the superfluid density vanishes and discuss two important consequences,
namely that the phonon collective mode has quadratic rather than the
expected \cite{jpephonon} linear dispersion and that long-wavelength
instabilities appear when $\Delta _{V}\neq 0$. We conclude with some
speculations on the experimental implications of our findings.

\noindent \emph{Octet mean-field theory for unbalanced bilayers}--- Bilayer
graphene's $N=0$ Landau-level octet is the direct product of three $S=1/2$
doublets: real spin and {\em which-layer} 
pseudospin as
in a semiconductor bilayer, and the Landau-level ($n=0,1$) pseudospin which
is responsible for the new physics we discuss in this paper. The octet
degeneracy is lifted by the Zeeman coupling $\Delta _{Z}$, assumed here to
maximize spin-polarization, and by $\Delta _{V}$
which is defined so that it favors top layer occupation when positive. In
graphene bilayers, $\Delta _{V}$ also drives \cite%
{graphenebilayertheory,ourprl} a small splitting of the Landau-level
pseudospin that plays a key role in this paper: $\Delta _{LL}=\alpha \Delta
_{V}\hbar \omega _{c}/\gamma _{1}$ where $\hbar \omega _{c}=2.14\,B[\mathrm{%
Tesla]}$ meV, $\gamma _{1}\sim 400$ meV is the interlayer hopping energy,
and $\alpha =+1(-1)$ for top(bottom) layers. The fact that LL $1$ has a
higher energy than LL $0$ in the top layer and a lower energy in the bottom
layer will prove to be important.

The octet mean-field Hamiltonian can be separated into single-particle and exchange contributions: 
\begin{eqnarray}
H_{HF} &=&E_{\alpha ,n,s}\rho _{\alpha ,s,n;\alpha ,s,n}  \label{hfham} \\
&&-\frac{1}{g}X_{n_{1},n_{4},n_{3},n_{2}}^{\left( \alpha ,\alpha \right)
}\left( 0\right) \left\langle \rho _{s,n_{1};r,n_{2}}^{\alpha ,\alpha
}\right\rangle \rho _{r,n_{3};s,n_{4}}^{\alpha ,\alpha }  \nonumber \\
&&-\frac{1}{g}X_{n_{1},n_{4},n_{3},n_{2}}^{\left( \alpha ,\overline{\alpha }%
\right) }\left( 0\right) \left\langle \rho _{s,n_{1};r,n_{2}}^{\alpha ,%
\overline{\alpha }}\right\rangle \rho _{r,n_{3};s,n_{4}}^{\overline{\alpha }%
,\alpha },  \nonumber
\end{eqnarray}%
(repeated indices are summed over) 
where the single particle energy (which includes the Hartree capacitive term) is 
\begin{equation}
E_{\alpha ,s,n}=-s\frac{\Delta _{Z}}{2}-\alpha \frac{\Delta _{V}}{2}+\alpha
n\Delta _{LL}+\frac{d}{l_{B}}\left( \frac{\nu }{2}-\nu _{\overline{\alpha }%
}\right) ,
\end{equation}%
and the exchange interactions 
\begin{eqnarray}
X_{n_{1},n_{2},n_{3},n_{4}}^{\left( \alpha ,\beta \right) }\left( \mathbf{q}%
\right) &=&\int \frac{d\mathbf{p}l_{B}^{2}}{2\pi }\frac{e^{-pd\left(
1-\delta _{\alpha ,\beta }\right) }}{pl_{B}}  \label{exchange} \\
&&\times F_{n_{1},n_{2}}\left( \mathbf{p}\right) F_{n_{3},n_{4}}\left( -%
\mathbf{p}\right) e^{i\mathbf{q}\times \mathbf{p}l_{B}^{2}}.  \nonumber
\end{eqnarray}%
In these equations, $g$ is the Landau level degeneracy, $d=3.337$\AA\ is the
interlayer separation, $l_{B}$ is the magnetic length, $\Delta _{Z}=g\mu
_{B}B$ is the Zeeman coupling, $n=0,1$ are LL indices, $\alpha ,\beta =t,b$
for top(bottom) layers, $r,s=1(-1)$ for up(down) spins and $\overline{t}=b,%
\overline{b}=t.$ All energies are in units of $e^{2}/\varepsilon l_{B}$. The
average value of the operators $\rho _{s,n_{1};t,n_{2}}^{\alpha ,\beta }=\sum_{X}c_{\alpha
,s,n_{1},X}^{\dagger }c_{\beta ,t,n_{2},X}$ (where $c_{\alpha
,s,n,X}^{\dagger }$ creates an electron with guiding-center $X$ in the
Landau gauge and layer $\alpha ,$ spin $s$ and Landau level character $n$)
must be determined self consistently by occupying the lowest energy
eigenvectors of $H_{HF}$. The form factors ($F_{00}(\mathbf{q}%
)=e^{-(ql_{B})^{2}/4}$, $F_{10}(\mathbf{q}%
)=(iq_{x}+q_{y})l_{B}e^{-(ql_{B})^{2}/4}/\sqrt{2}=[{F}_{01}(-\mathbf{{q}%
)]^{\ast }}$ and $F_{11}(\mathbf{q})=(1-(ql_{B})^{2}/2)e^{(-ql_{B})^{2}/4}$)
reflect the character of the two different quantum cyclotron orbits. 

Although an infinitesimal $\Delta _{V}$ would be sufficient to produce
complete layer polarization in a non-interacting system, a finite value is
required once interactions are accounted for. Spontaneous interlayer phase
coherence arises because it is able to lower energy by inducing a gap at the
Fermi level even when both layers are partially occupied. When the layer
index is viewed as a pseudospin \cite{coherencerefs}, the phase-coherent
state corresponds to an $\hat{x}-\hat{y}$ easy-plane ferromagnet and the
interlayer phase difference corresponds to the azimuthal magnetization
orientation. In this language $\Delta _{V}$ is a hard-direction external
field which gradually tilts the magnetization direction toward the $\hat{z}$
direction. We find that for $\Delta _{V}\geq \Delta _{V}^{\ast }\left(
B\right) ,$ with $\Delta _{V}^{\ast }\left( B\right) /(e^{2}/\varepsilon
l_{B})\approx 0.001$ at $B=10$T, the system exhibits full charge imbalance
with all electrons (all minority spin electrons in the 
$\nu_{T}=1$ case)  occupying one layer.  The critical $\Delta
_{V} $ is given by%
\begin{equation}
\Delta _{V}^{\ast }=\frac{e^{2}}{\varepsilon l_{B}}\left( \frac{d}{l_{B}}-%
\sqrt{\frac{\pi }{2}}\left[ 1-e^{d^{2}/2l_{B}^2} {\rm Erfc}\left[ \frac{d}{\sqrt{2}l_{B}}%
\right] \right] \right) .
\end{equation}%
In the graphene bilayer case, the $\Delta _{V}$ sufficient to achieve full
pseudospin polarization is reduced compared to the semiconductor case
because the layers are close together. For $\Delta _{V}\leq \Delta
_{V}^{\ast }$, the charge-unbalanced mean-field ground state consists of a
full Landau level of states \cite{ourprl,otherstatecaveat,yogesh} which
share partially polarized layer and $n=0$ Landau-level pseudospinors: 
\begin{equation}
|\Psi _{GS}\rangle =\prod_{X}\left[ \cos \left( \frac{\theta _{V}}{2}\right)
\,c_{t,0,X}^{\dagger }+e^{i\varphi }\sin \left( \frac{\theta _{V}}{2}\right)
\,c_{b,0,X}^{\dagger }\,\right] \left\vert 0\right\rangle ,  \label{eq:gs}
\end{equation}%
where we have dropped the irrelevant spin degree-of-freedom. In Eq.(\ref%
{eq:gs}), $\varphi $ is the spontaneous interlayer phase and $\cos \left(
\theta _{V}\right) =\Delta _{V}/\Delta _{V}^{\ast }\left( B\right) .$

\noindent \emph{Vanishing superfluid density at zero bias}--- The energy
cost of phase gradients is the key property of any superfluid. In normal
superfluids, including semiconductor bilayer exciton superfluids, the
leading term in a phase-gradient expansion of the energy-density is
proportional to $|\nabla \varphi |^{2}$. We now show that the coefficient
which specifies the size of this term, often referred to as the superfluid
density, is zero in bilayer graphene exciton condensates at zero bias. We
proceed by explicitly constructing a pseudospin wave state in which $\nabla \varphi =q\hat{x}
$ is constant: 
\begin{equation}
|\Psi _{q}^{PSW}\rangle =\prod_{X}\left[ \cos \left( \frac{\theta _{V}}{2}%
\right) c_{t,0,X}^{\dagger }+e^{iqX}\sin \left( \frac{\theta _{V}}{2}\right)
\Lambda _{q,X}^{\dag }\right] \left\vert 0\right\rangle ,
\label{eq:gradstate}
\end{equation}%
where $\Lambda _{q,X}^{\dag }=u_{q}c_{b,0,X}^{\dagger
}+v_{q}c_{b,1,X}^{\dagger }$ and $u_{q}^{2}+v_{q}^{2}=1$. In this
wavefunction the factor $\exp (iqX)$ is responsible for the phase gradient.
The term proportional to $v_{q}$ in Eq. (\ref{eq:gradstate}) allows for the
crucial possibility, unique to bilayer graphene, of combining the phase
gradient with an admixture of the $n=1$ wavefunction. 

In mean-field theory the energy is
proportional to the square of the density-matrix.
We now show that it is possible to choose $v_{q}$ so that the density matrix
is unchanged to first order in $q$ and that 
the superfluid density vanishes as a consequence. Since $v_{q}$ must vanish for $%
q\rightarrow 0$ to reproduce $|\Psi _{0}\rangle $, we have $v_{q}\sim q$ and 
$u_{q}\sim 1$ to this order. It is then easy to see that the density-matrix
component within each layer is unchanged to leading order in $q$. In quantum
Hall bilayer exciton condensates, the superfluid density is due to reduced
interlayer exchange energy in the presence of a phase gradient\cite%
{coherencerefs,abolfath}, and hence to changes in the interlayer density
matrix, $\rho _{tb}\left( \mathbf{r},\mathbf{r}^{\prime }\right)
=\left\langle \Psi _{t}^{\dag }\left( \mathbf{r}\right) \Psi _{b}\left( 
\mathbf{r}^{\prime }\right) \right\rangle $ where the field operator $\Psi
_{t\left( b\right) }^{\dag }\left( \mathbf{r}\right) = 1/\sqrt{L_{y}}
\sum_{n=0,1}\sum_{X}\phi _{n}^{\ast }\left( x-X\right)
e^{-iXy/l_{B}^{2}}c_{t\left( b\right) ,n,X}^{\dag }$ with $\phi _{n}(x)$ the
harmonic oscillator state with orbital Landau character $n$. At zero bias,
Eq. (\ref{eq:gs}) with $\varphi =0$ gives $\rho _{tb}^{GS}\left( \mathbf{r},
\mathbf{r}^{\prime }\right) = 1/(2L_{y})\sum_{X} \phi _{0}^{\ast }\left(
x-X\right) \phi _{0}\left( x^{\prime }-X\right) e^{-iX\left( y-y^{\prime
}\right) /\ell_{B} ^{2}}$ while Eq. (\ref{eq:gradstate}) implies%
\begin{eqnarray}
\nonumber
\rho _{tb}^{PSW}(\mathbf{r},\mathbf{{r^{\prime }}}) &=&\frac{1}{2L_{y}}%
\sum_{X}\phi _{0}^{\ast }\left( x-X\right) e^{-iX\left( y-y^{\prime }\right)
/l_{B}^{2}}e^{iqX} \\
&&\times \left[ u_{q}\phi _{0}\left( x^{\prime }-X\right) +v_{q}\phi
_{1}\left( x^{\prime }-X\right) \right]  
\end{eqnarray}
Since $\phi _{1}(x)=\sqrt{2}(x/l_{B})\phi _{0}(x)$ and $q(x'-X)$ is small
because of the localized oscillator wavefunctions, it follows that $\rho
_{tb}^{GS}(\mathbf{r},\mathbf{r}^{\prime })$ is altered only by an irrelevant phase factor to first 
order in $q$ if we choose $u_{q}=1,v_{q}= iql_{B}/\sqrt{2}$.  The 
leading change in $|\rho_{tb}|$ is therefore proportional to $q^2$
and the leading energy change proportional to $q^4$. 

\begin{figure}[t]
\begin{center}
\includegraphics[clip,width=3.375in,height=2.25in]{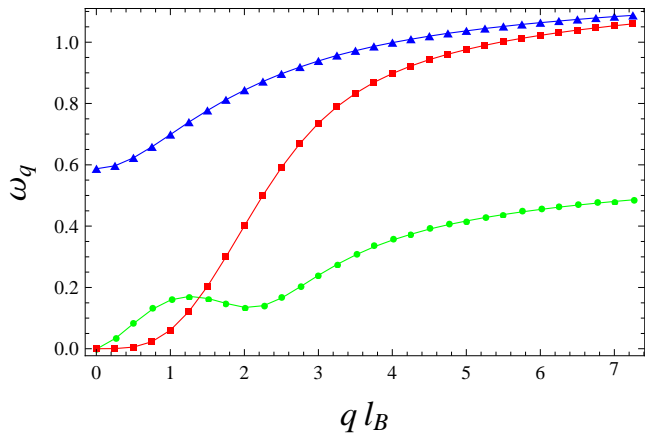}
\end{center}
\caption{(Color online) Collective mode dispersion $\protect\omega_{q}$ for
balanced Octet QHF in units of interaction strength $e^2/\protect\varepsilon %
l_{B} = 11.2 (B[T])^{1/2}$ meV as a function of $ql_{B}$ at a magnetic field
of $B = 10$ Tesla. The green (circle) curve shows the inversion symmetric $%
n=0$ to $n=1$ transition mode with $\lim_{q \to 0} \protect\omega_{1} \sim
q^{3/2}$ dispersion which is unrelated to superfluid properties. The red
(box) curve shows the inversion asymmetric superfluid Goldstone mode with
the anomalous long-wavelength dispersion $\lim_{q\to 0} \protect\omega_{2}
\sim q^{2}$. The blue (triangle) curve is a gapped mode not essential for our discussion 
(see text for details).}
\label{fig2}
\end{figure}

\noindent \emph{Unusual collective mode dispersion}--- We now consider
collective excitations of the bilayer quantum Hall exciton condensate at $%
\nu _{T}=-3,1$, first for balanced ($\Delta _{V}=0$) bilayers. As discussed
above, the ground state is a full Landau level with shared layer symmetric $%
n=0$ pseudospin orbitals $|+,0\rangle =\left( |t,0\rangle +|b,0\rangle
\right) /\sqrt{2}$. The pseudospinor fluctuations that are quantized in the
system's collective modes mix in components from the three orthogonal
pseudospin states: $|+,1\rangle $ also symmetric in layer indices and $%
|-,0\rangle $ and $|-,1\rangle $ antisymmetric in layer indices. Fig. \ref%
{fig2} shows the collective mode spectrum calculated in the time-dependent
mean-field theory \cite{coteprb}. The inversion symmetric mode (transition $%
|+,0\rangle \rightarrow |+,1\rangle $) is unrelated to superfluidity and has
been discussed previously \cite{ourprl}. For $q=0$ the two asymmetric modes
correspond respectively to global rotations of the $|+,0\rangle $
pseudospinor toward the states $|-,0\rangle $ and $|-,1\rangle $
respectively. The later rotation costs a finite energy because the exchange
energy in $n=1$ is smaller than in $n=0$.

One physically transparent way of performing these collective mode
calculations is to construct a fluctuation action in which each transition
has canonically conjugate density $\rho $ and phase $\varphi $ components
corresponding to the real and imaginary parts of the final state component
in the fluctuating spinor. 
The fluctuation action 
\begin{equation}
\mathcal{S}[\mathbf{\rho },\mathbf{\varphi }]=\mathcal{S}_{B}[\mathbf{\rho },%
\mathbf{\varphi }]-\int d\omega \int d^{2} q \;\mathcal{E}[\mathbf{\rho },%
\mathbf{\varphi }],  \label{action}
\end{equation}%
contains a Berry phase term $\mathcal{S}_{B}$~\cite{auerbach} 
\begin{equation}
\mathcal{S}_{B}[\mathbf{\rho },\mathbf{\varphi }]=\int d\omega \int d^{2}q %
\left[ \frac{1}{2}\mathbf{\rho }_{-\mathbf{q}}^{\dag }\mathcal{D}\mathbf{%
\rho }_{\mathbf{q}}-\mathbf{\varphi }_{-\mathbf{q}}^{\dag }\mathcal{D}%
^{\dagger }\mathbf{\varphi }_{\mathbf{q}}\right] ,
\end{equation}%
where $\mathbf{\rho }_{\mathbf{q}}=(\rho _{1,\mathbf{q}},\rho _{2,\mathbf{q}%
},\rho _{3,\mathbf{q}})$ and $\mathcal{D}=-i\omega \mathbb{I}_{3\times 3}$,
and an energy functional $\mathcal{E}[\mathbf{\rho },\mathbf{\varphi }]$
closely related to the discussion in the preceding section 
\begin{equation}
\mathcal{E}[\mathbf{\rho },\mathbf{\varphi }]=\frac{1}{2}\left[ \mathbf{\rho 
}_{-\mathbf{q}}^{\dag }\Lambda _{\rho }(q)\mathbf{\rho }_{\mathbf{q}}+%
\mathbf{\varphi }_{-\mathbf{q}}^{\dag }\Lambda _{\varphi }(q)\mathbf{\varphi 
}_{\mathbf{q}}\right] .
\end{equation}%
Here, $\Lambda _{\rho }(q)$ and $\Lambda _{\varphi }(q)$ capture the energy
cost of small pseudospinor fluctuations and can be evaluated explicitly.
Because $\rho $ fluctuations change the charge distribution in the system
they remain finite for $q\rightarrow 0$ and do not play an essential role in
our discussion. At long wavelengths, we find that the inversion asymmetric
block in $\Lambda _{\varphi }$ has the form 
\begin{equation}
\left( 
\begin{array}{cc}
\frac{X_{1001}^{\left( +-\right) }(0)}{2}q^{2}l_{B}^{2}+\cdots  & i\frac{%
X_{1001}^{\left( +-\right) }(0)}{\sqrt{2}}ql_{B}+\cdots  \\ 
-i\frac{X_{1001}^{\left( +-\right) }(0)}{\sqrt{2}}ql_{B}+\cdots  & 
X_{1001}^{\left( +-\right) }(0)+\cdots 
\end{array}%
\right) ,  \label{balancedredmat}
\end{equation}%
with \textquotedblleft $\cdots $\textquotedblright\ representing terms higher
order in $ql_{B}$. 

In a semiconductor bilayer only $n=0$ phase fluctuations are possible without paying a 
kinetic energy penalty.  In the strong magnetic field limit the superfluid density is 
therefore proportional to the interlayer
exchange-interaction constant $X_{1001}^{\left( +-\right) }$.
For the bilayer graphene octet, however, the energy cost of phase variation is
proportional to the smallest eigenvalue of the matrix in Eq.(\ref{balancedredmat}) and this 
has the $q^4$-long wavelength dependence anticipated above. 
Indeed the eigenvector of this quadratic form captures the orbital character of the Goldstone mode 
\begin{equation}
|GM\rangle =|-,0\rangle + i\frac{ql_{B}}{\sqrt{2}}|-,1\rangle ,
\end{equation}%
corresponding to the $v_{q}/u_{q}$ ratio in that analysis. We find that the
leading small $q$ behavior for the energy functional is 
\begin{equation}
\mathcal{E}_{-}[\varphi _{-}]=\beta (q l_{B})^{4}\varphi _{-}^{2}+\cdots ,
\end{equation}%
where $\beta =0.093(e^{2}/\varepsilon l_{B})$ at $10$T and $\varphi _{-}$ is
the amplitude of the lowest-energy eigenmode of the phase matrix $\Lambda
_{\varphi }(q)$. 
Because of the conjugate relationship between $%
\rho $ and $\varphi $ fluctuations, the collective mode energy is
proportional to the square root of the eigenvalues of $\Lambda _{\varphi }$
and $\Lambda_{\rho}$ so that the quadratic Goldstone mode dispersion simply 
signals the vanishing superfluid density.\newline
\emph{Long-wavelength instability at finite bias}--- The $q^{4}$ interaction
energy cost of inter-layer phase gradients holds for balanced and unbalanced
bilayers. In unbalanced layers, however, the Landau level splitting $\Delta
_{LL}$ means that the single-particle energy is lowered by transitions to
the $n=1$ Landau level in the bottom layer. For unbalanced layers we find
that the phase fluctuation action goes to 
\begin{equation}
\left( 
\begin{array}{cc}
\frac{X_{1001}^{\left( +-\right) }(0)}{2}q^{2}l_{B}^{2}+\cdots  & i\frac{%
X_{1001}^{\left( +-\right) }(0)}{\sqrt{2}}ql_{B}+\cdots  \\ 
-i\frac{X_{1001}^{\left( +-\right) }(0)}{\sqrt{2}}ql_{B}+\cdots  & 
X_{1001}^{\left( +-\right) }(0)-\cos \left( \theta _{V}\right) \Delta
_{LL}+\cdots 
\end{array}%
\right) .  \label{balancedredmat2}
\end{equation}%
The energy cost of phase fluctuations is again proportional to the smallest
eigenvalue of the phase matrix,
\begin{equation}
\mathcal{E}_{-}[\varphi _{-}]=-\frac{1}{2}\cos \left( \theta _{V}\right)
\Delta _{LL}(q l_{B})^{2}\varphi _{-}^{2}+\beta (q l_{B})^{4}\varphi _{-}^{2}+\cdots ,
\label{unbalancedenergy}
\end{equation}%
%
%
which is negative at small $q$, indicating that a uniform condensate is unstable when $\Delta_{LL} \ne 0$.
The energy functional in 
Eq.(\ref{unbalancedenergy}) reduces to the classical Swift-Hohenberg (SH)
model~\cite{brazovskiiswifth} Hamiltonian:   
\begin{equation}
\mathcal{E}[\varphi_{-} ]=\left[ \frac{\Delta _{0}}{2}+\frac{\xi _{0}^{4}}{2}%
(\nabla^{2} + q_{0}^{2})^{2}\right] \varphi_{-} ^{2}+\frac{\lambda }{4!}\varphi_{-}^{4},
\end{equation}%
when we set $\Delta _{0}=-(\cos \left( \theta _{V}\right) \Delta _{LL})^{2}/8\beta $%
, $\xi_{0}^{2}=\sqrt{2\beta} l_{B}^{2}$ and $(q_{0} l_{B})^2= \cos \left( \theta
_{V}\right) \Delta _{LL}/4\beta $. 
Using the detailed microscopic form of the mean-field energy functional,
we estimate that $\lambda \sim
(q_{0}^{4} l_{B}^{2}) (e^{2}/\varepsilon l_{B})$. The SH model exhibits a
fluctuation-induced first order smectic-isotropic phase transition from a
stripe ordered phase with $\langle \varphi_{-} \rangle =A\cos (\mathbf{q}_{0}\cdot 
\mathbf{r})$ to a disordered phase with $\langle \varphi_{-} \rangle =0$. Following
the self-consistent Hartree approximation analysis\cite{brazovskiiswifth} of the Swift-Hohenberg model, 
for $\Delta_{V} < \Delta_{V}^{\ast}$ we estimate the transition temperature to be  
\begin{equation}
k_{B}T_{c}=\frac{4\xi _{0}^{2}}{2.03\lambda}|\Delta _{0}|^{3/2}=1.97 
\frac{\beta}{e^2/\varepsilon l_{B}} \left( \frac{\hslash \omega _{c}}{\gamma _{1}}
\Delta _{V}^{\ast }\right) \left( 
\frac{\Delta _{V}}{\Delta _{V}^{\ast }}\right) ^{2}.
\end{equation}
which is typically below $10 {\rm mK}$. 
%
%

\noindent \emph{Discussion}--- 
We identify the $T>T_{c}$
phase with a normal quantum Hall ferromagnet and the $T<T_{c}$ phase with a
quantum Hall smectic state which should exhibit
anisotropic transport properties.
For $T > T_{c}$ we expect properties similar to those observed in 
semiconductor bilayers\cite{transportrefs} except that the  
coherent interlayer tunneling processes, which plays a key role 
in tunneling experiments, should be essentially absent.  In graphene 
bilayers therefore, spontaneous coherence is likely to be 
most conveniently manifested by strongly enhanced Hall drag.
Finally we note that trigonal warping terms, which we have neglected, will 
reduce $\Delta _{LL}$\cite{ourprl} by at most $5\%$ for magnetic
fields of interest, that correlation effects we have not considered 
could contribute positively or negatively to the superfluid density, and 
that the small superfluid densities in this system might lead to 
important thermal fluctuation effects beyond those considered here.
This work was supported by the NSF under grant DMR-0606489 (AHM) a grant by NSERC (RC) and 
the State of Florida (YB). 

\noindent %
%

\end{document}